\documentclass[prb,preprint]{revtex4}
\usepackage[pdftex]{graphicx}
\pdfoutput=1

\newcommand{\ignore}[1]{}

\begin{document}

\title
{Macroscopic nucleation phenomena in continuum media with long-range interactions}

\author{Masamichi Nishino$^{1,2,7}$}  
\author{Cristian Enachescu$^{3}$}
\author{Seiji Miyashita$^{4,7}$}
\author{Per Arne Rikvold$^{5}$}
\author{Kamel Boukheddaden$^{6}$}
\author{Fran\c cois Varret$^{6}$}
\affiliation{$^{1}${\it Computational Materials Science Center, National Institute}  
for Materials Science, Tsukuba, Ibaraki 305-0047, Japan \\
$^{2}${\it Institute for Solid State Physics, the University of Tokyo, Kashiwa, Japan} \\
$^{3}${Department of Physics, Alexandru Ioan Cuza University, Iasi, Romania} \\
$^{4}${\it Department of Physics, Graduate School of Science,} The University of Tokyo, Bunkyo-Ku, Tokyo, Japan  \\
$^{5}$ {\it Department of Physics, Florida State University, Tallahassee, 
Florida 32306-4350, USA} \\
$^{6}${\it  Groupe d'Etudes de la Mati\`{e}re Condens\'{e}e, CNRS-Universit\'{e}} de Versailles/St. Quentin en Yvelines, 45 Avenue des Etats Unis, F78035 Versailles Cedex, France \\
$^{7}${\it CREST, JST, 4-1-8 Honcho Kawaguchi, Saitama, 332-0012, Japan}
}

\begin{abstract}

{\bf 

Nucleation, commonly associated with discontinuous transformations
between metastable and stable phases, is crucial in fields as diverse as
atmospheric science and nanoscale electronics.  Traditionally, it is
considered a microscopic process (at 
most nano-meter), implying the formation of a 
microscopic nucleus of the stable phase. 
Here we show for the first time, that
considering long-range interactions mediated by elastic distortions,
nucleation can be a macroscopic process, with the size of the critical
nucleus proportional to the total system size. 
This provides a new concept of ``macroscopic barrier-crossing nucleation". 
We demonstrate the effect in molecular dynamics simulations of a model spin-crossover system with two molecular states of different sizes, causing elastic distortions. 
}
 
\end{abstract}

\maketitle

\begin{figure}[t]
  \begin{minipage}{0.35\hsize}
  \begin{center}
    \includegraphics[width=35.5mm]{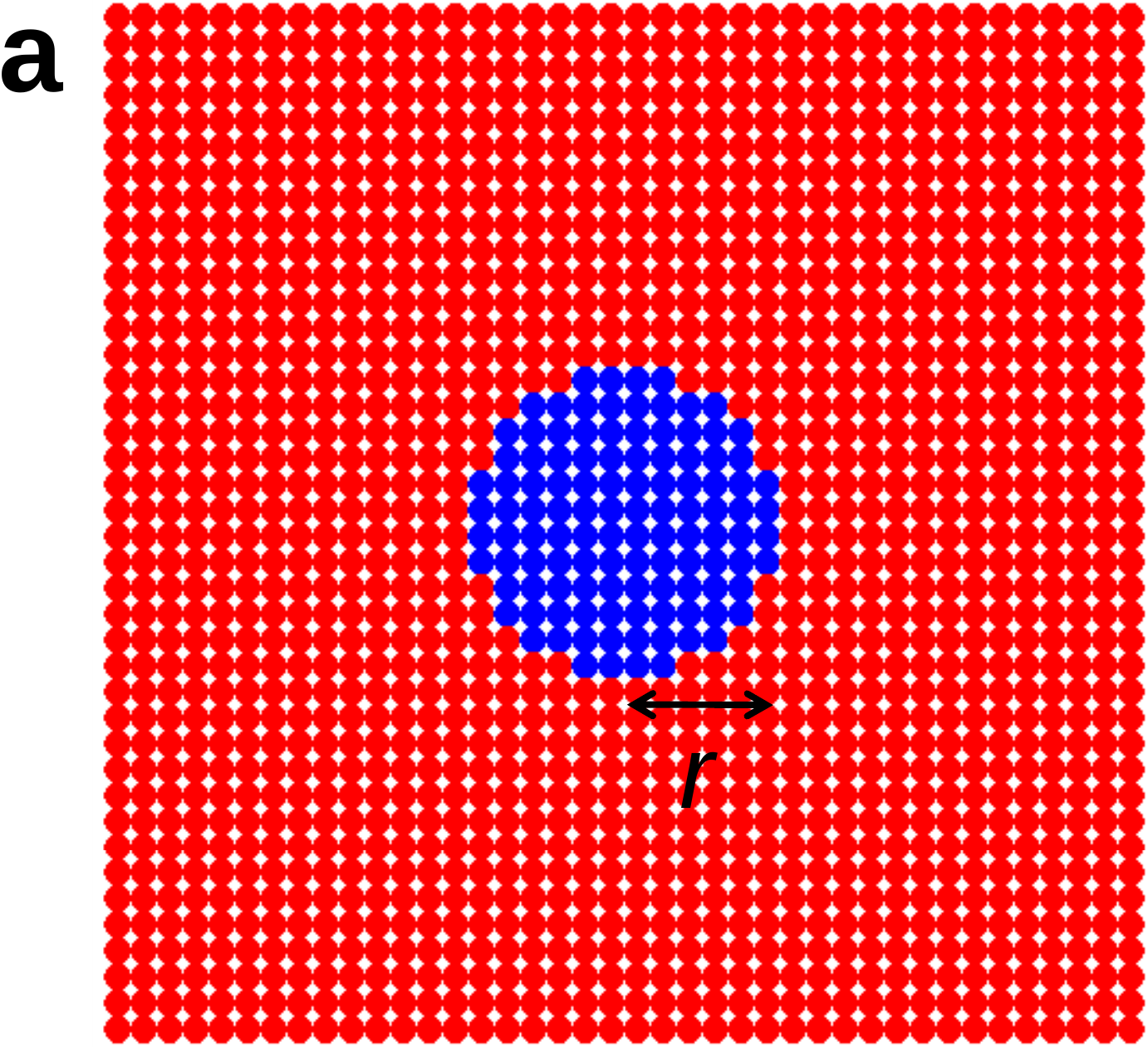}
  \end{center}
 \end{minipage}
   \begin{minipage}{0.35\hsize}
   \begin{center}
   \includegraphics[width=35.5mm]{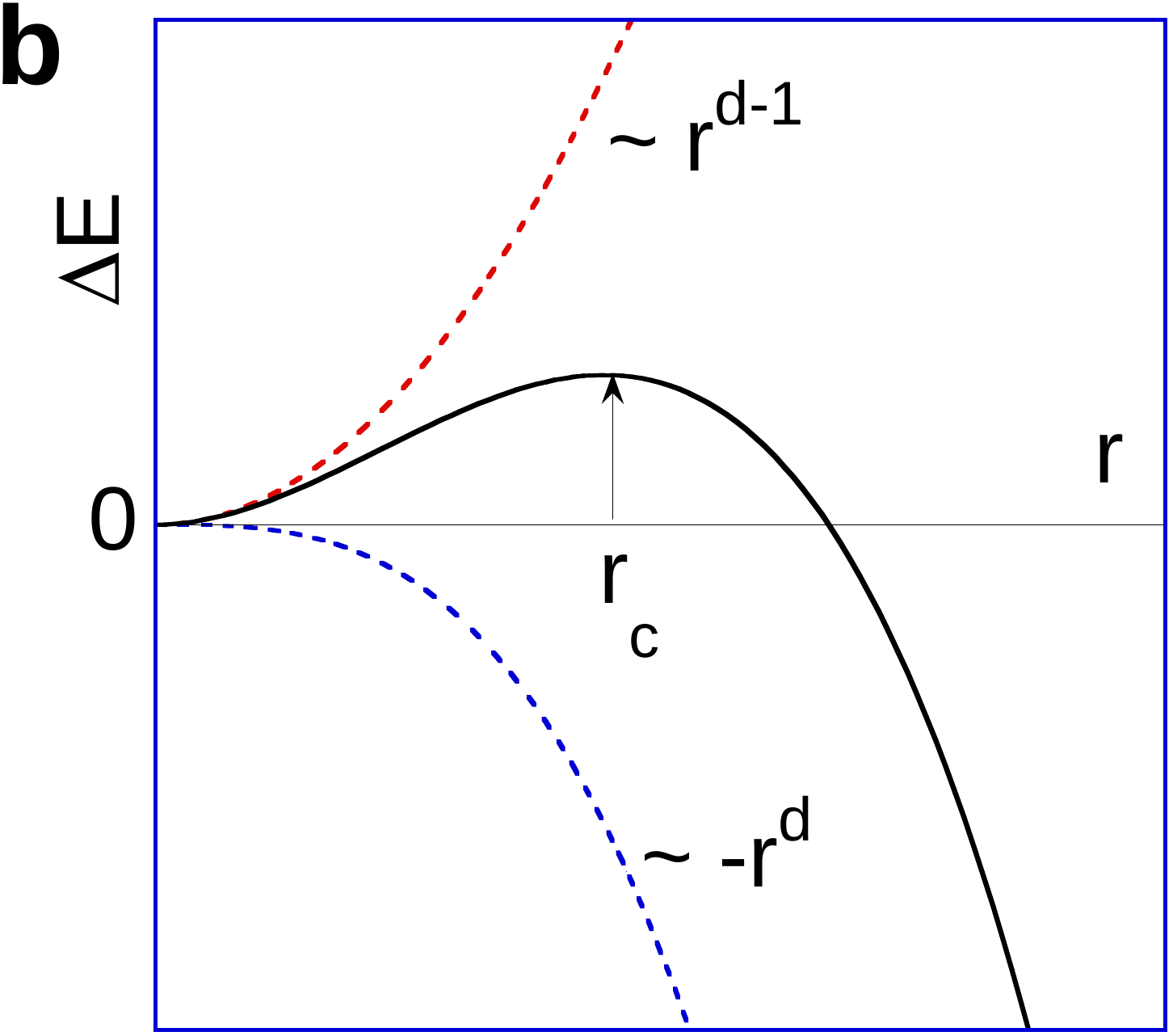}
  \end{center}
 \end{minipage}

\vspace*{1.5cm}

\centerline{\includegraphics[clip,width=9cm]{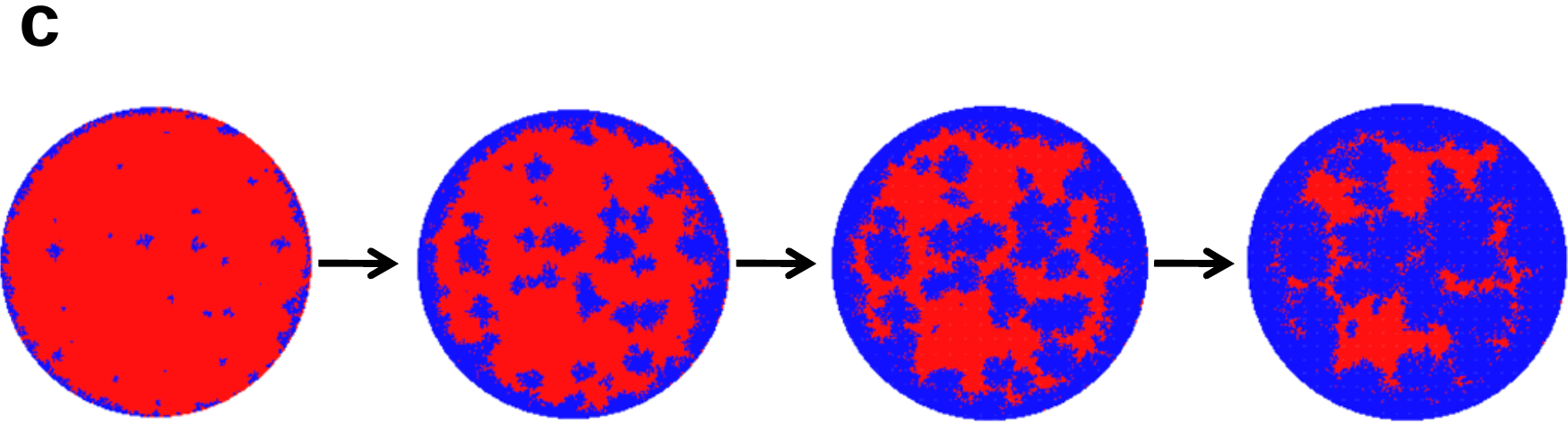}}
\caption{ {\bf Nucleation and domain formation for short-range interaction systems.} 
{\bf a}, A schematic example of a droplet of a short-range magnetic 
interaction system. The blue circle with radius $r$ is a cluster of down-spin molecules in an up-spin phase (red part). 
{\bf b}, Microscopic competition of the surface and bulk free energies of a 
droplet. The surface free energy is an increasing function of the radius of 
the droplet ($r$) (upper broken line) and the bulk free energy is a decreasing function of $r$  (lower broken line). 
The solid line is the sum of these energies ($\Delta E$). 
The critical nucleus with the radius $r_{\rm c}$ gives the maximum of the total free energy. 
{\bf c}, Snapshots of nucleation for a short-range interaction model (Ising model). Red and blue denote up and down spins, respectively. 
} 
\label{Fig_classical}
\end{figure}

Nucleation is a barrier-crossing process,~\cite{Abraham_book,Kurz_book,Kelton} 
in which a metastable phase decays via a critical nucleus for which the 
increase in surface free energy is compensated by the bulk energy decrease.~\cite{Abraham_book,Kurz_book,Kelton,Schmelzer,Mo,Tersoff,Cacciuto,Shin,Zhang,Lindenmeyer}
If the cluster becomes bigger than the critical size, it grows, 
while if smaller, it shrinks. 
The size of a critical nucleus is determined by microscopic competition 
between the 
surface and bulk free energies of a microscopic cluster, and thus the size of 
the critical nucleus is microscopic~\cite{Kelton,Zhang,Shin,Cacciuto} (see Fig.\ref{Fig_classical}a and b). 
To be precise, this situation is realized in short-range interaction systems, 
where separation of the energy between the bulk and surface is allowed. 
Nucleation theories~\cite{Abraham_book,Kurz_book,Kelton,Schmelzer,Lindenmeyer} 
have been based on this idea, and 
so far only microscopic nucleation is known. 

Consider a typical short-range interaction system at low temperatures: the 
$d$-dimensional Ising model~\cite{Baxter} defined by the Hamiltonian, 
${\cal H}=J\sum_{i,j} \sigma_i \sigma_j -h \sum_{i} \sigma_i$, where 
$\sigma=\pm 1$ (up and down spins). 
The free-energy barrier for a droplet with radius $r$ is 
$\Delta E=-C_{\rm b} h r^{d}+ C_{\rm s} J r^{d-1}$. 
Here $C_{\rm b}r^{d}$ is proportional to the volume of the droplet 
and $C_{\rm s}r^{d-1}$ is proportional to the area of the phase boundary. 
As depicted in Fig.~\ref{Fig_classical}b, the radius of the critical droplet is 
given when the droplet has the maximum excess free energy as 
$r_{\rm c} =  \frac{C_{\rm s} (d-1)J}{C_{\rm b} d h}$. 
The critical radius $r_{\rm c}$ is independent of the system size. 
Figure~\ref{Fig_classical}c shows an example of nucleation in a circular 
system (open boundary conditions, OBC) for the Ising model ($d=2$). 
Nucleation takes place both in the bulk (inside) and at the boundary. 
Nucleation at the boundary is energetically more favorable, but 
when the system becomes larger (the bulk-to-boundary ratio becomes large), 
nucleation in the bulk becomes dominant (Supplementary 1). 

However, when the interaction is of long range, the nature of nucleation is 
different. 
Because it has been pointed out that an elastic interaction due to lattice 
distortion causes an effective long-range interaction,~\cite{Miyashita_elastic} 
the nucleation process in systems with elastic interactions (e.g., 
spin-crossover systems,~\cite{Gutlich_book,Konig,Kahn,Hauser}  martensitic 
systems~\cite{Porter,Bhattacharya,Richard,Tanaka} and Jahn-Teller systems~\cite
{Jahn,Goodenough,Englman,Bersuker}) should be investigated. 
In this work we present properties 
of the nucleation in 
long-range elastic interaction systems with OBC.

\begin{figure}
  \begin{minipage}{0.35\hsize}
  \begin{center}
    \includegraphics[width=42mm]{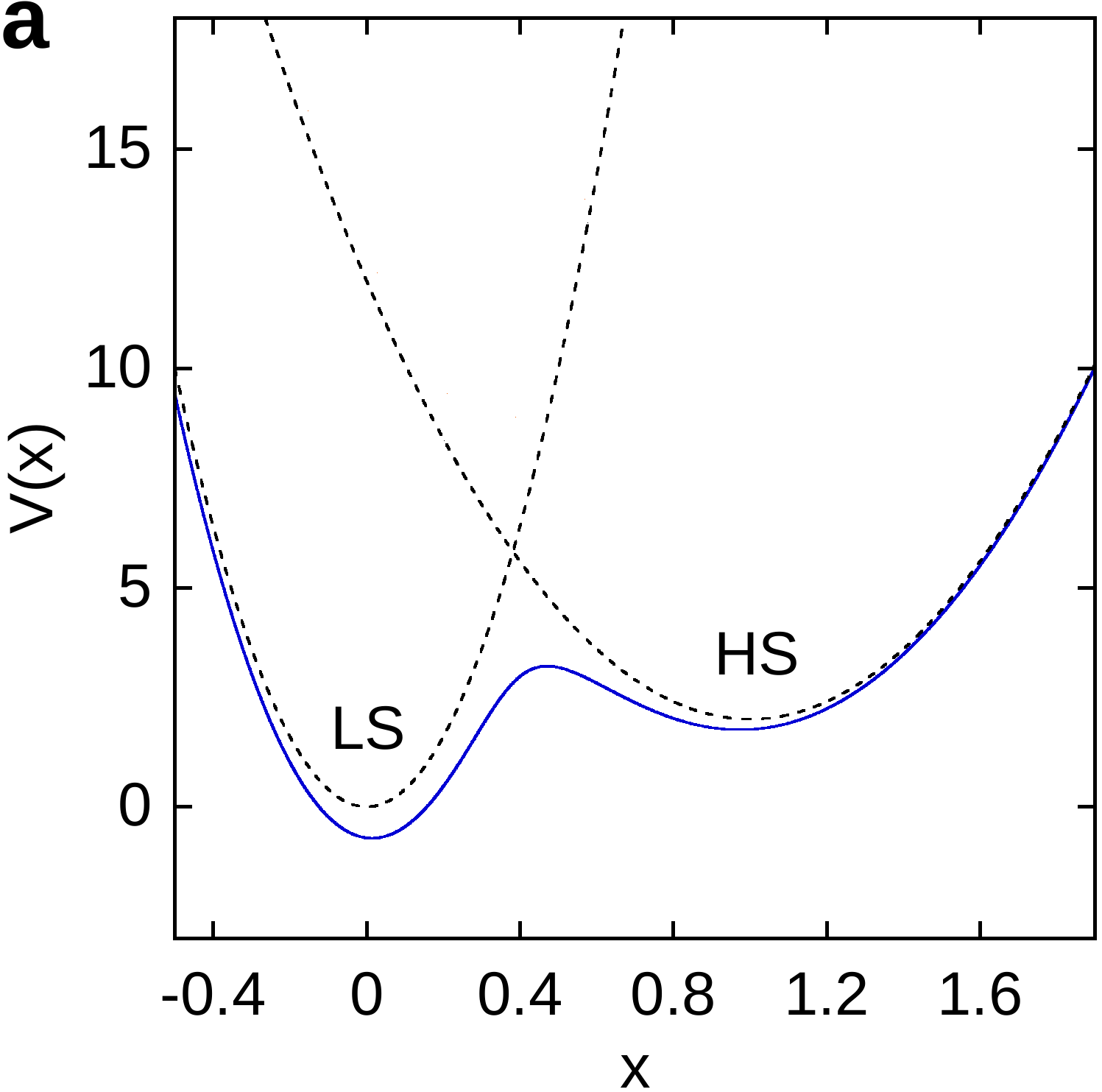}
  \end{center}
 \end{minipage}
   \begin{minipage}{0.30\hsize}
   \begin{center}
  \includegraphics[width=40mm]{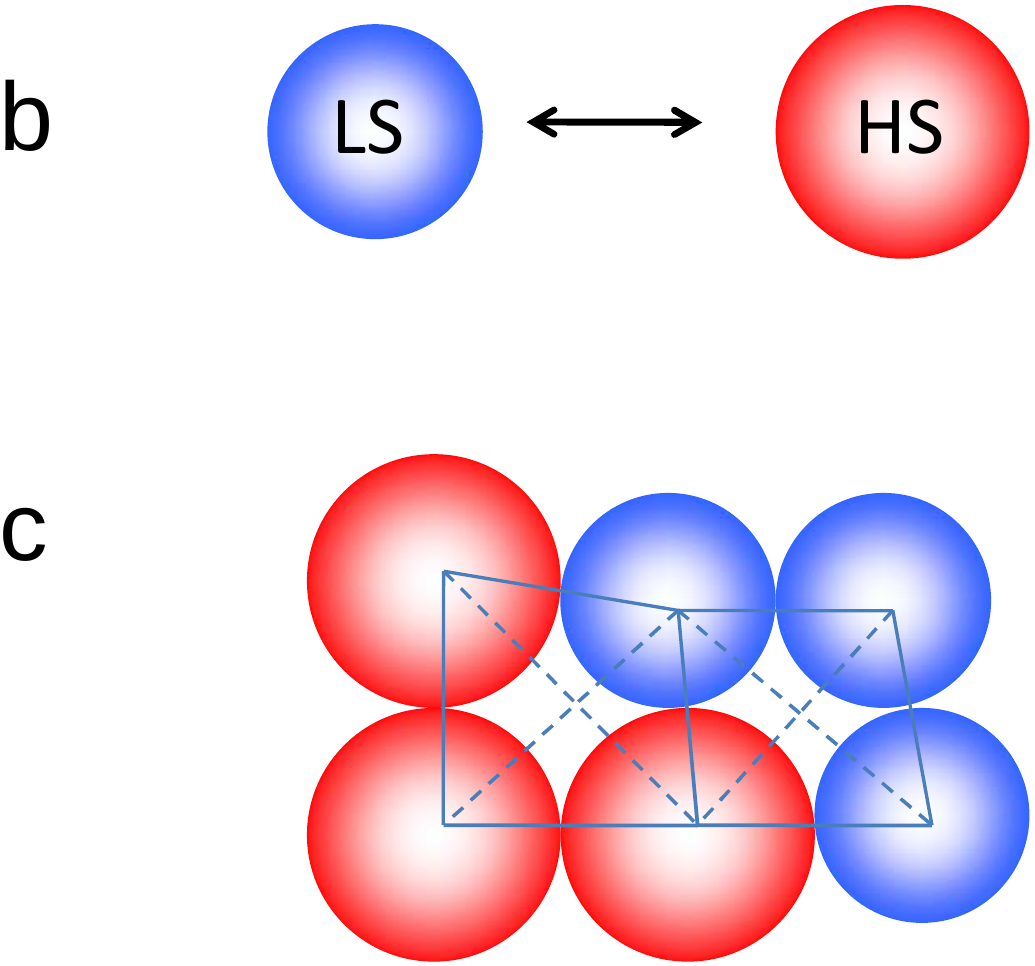}
  \end{center}
 \end{minipage}

\vspace*{3.0cm}
\centerline{\includegraphics[clip,width=11.4cm]{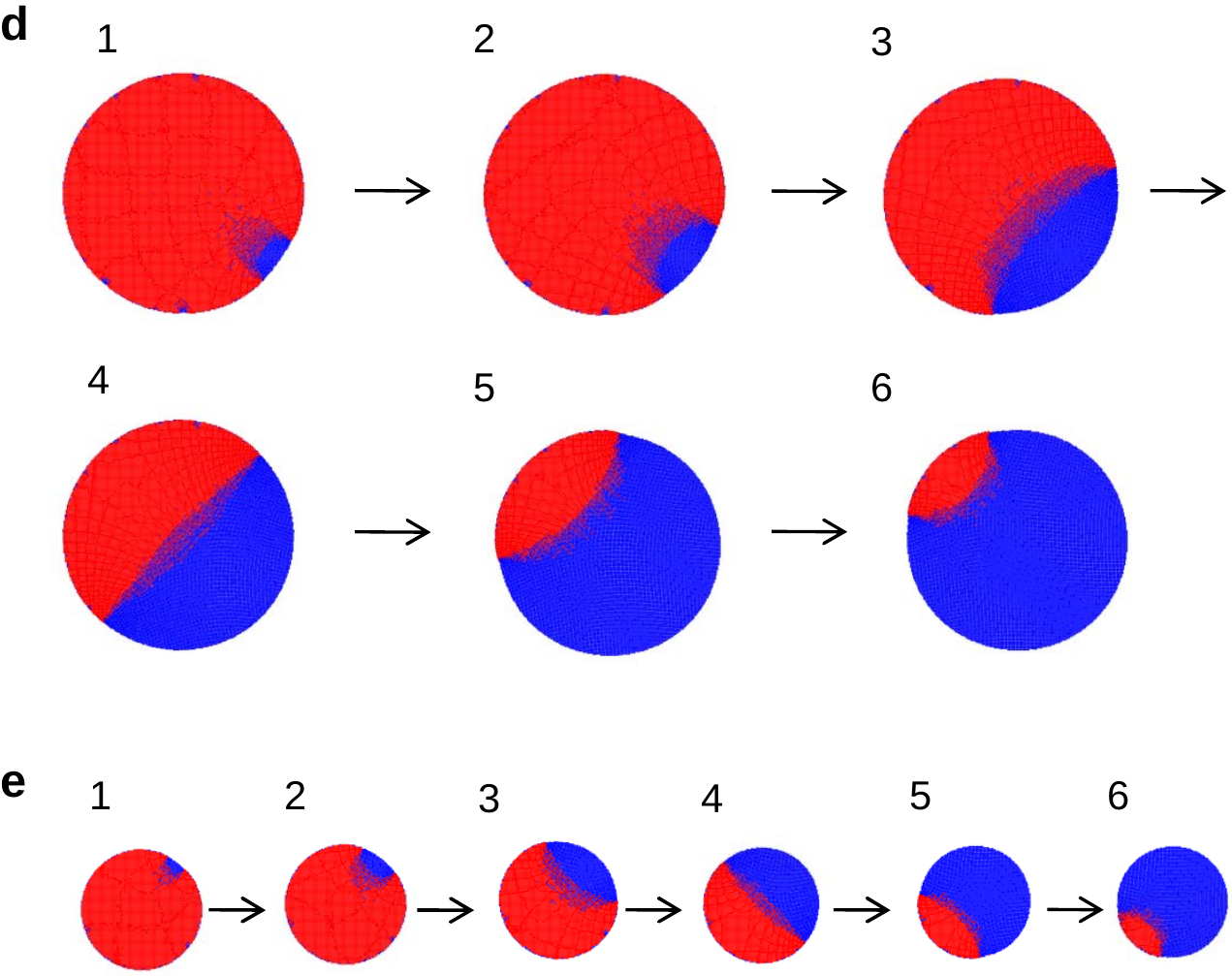} }
\caption{ {\bf Elastic interaction system and nucleation features.} 
{\bf a}, Intramolecular potential energy $V(x)$ shown by the
 solid (blue) curve. The dotted curves are low spin (LS) and high spin (HS) potential energies without quantum mixing. The curvature for the LS state is 4 times larger than for the HS state~\cite{Gutlich_book} in this work 
but other choices for the ratio of the curvatures, for example equal 
curvatures, do not change the essence of the results. 
The energy unit is 100$-$300 K for SC compounds~\cite{Gutlich_book} (Supplementary 2), and it is also the unit of the temperature $T$. 
{\bf b}, LS molecule (blue) and HS molecule (red). The HS molecule is larger in size.  
{\bf c}, Distortion due to the difference of the molecular sizes. 
{\bf d}, {\bf e}, Snapshots of the configuration during relaxation from the HS phase for ({\bf d}) $2R=200$ and ({\bf e}) $2R=100$. 
The value of $f_{\rm HS}$ is 0.95 for d1 and e1, 0.90 for d2 and e2, 
0.71 for d3 and e3, 0.50 for d4 and e4, 0.26 for d5 and e5, and 
0.15 for d6 and e6.
}
\label{Fig_model}
\end{figure}

In molecular crystals, e.g., transition-metal complexes, a
molecule often displays bistability in both its electronic state and 
molecular size (structure). External stimuli, e.g., change of temperature, 
pressure, photoirradiation, etc. change the molecular size. 
The distortion caused by the change of size induces an elastic interaction, 
which acts as an effective long-range interaction.~\cite{Miyashita_elastic,
Nishino_2010,Enachescu} 
Spin-crossover (SC) compounds are a 
typical example of the above situation 
(see Fig.\ref{Fig_model} a-c), 
where the low-spin (LS) and high-spin (HS) states are separated by an 
energy barrier, and the LS molecule is smaller than the HS one. Indeed, SC 
systems show a wide variety of phase transitions under external stimuli
.~\cite{Gutlich_book,Konig,Kahn,Hauser} 

In the present work we study the nucleation dynamics of circular ($d=2$) 
crystals of a long-range elastic interaction system and show that 
the nucleation is a barrier-crossing process. However,  
the size of the critical nucleus ($r_{\rm c}$) is proportional to the system 
size ($R$). Thus, a macroscopic nucleation 
mechanism is realized, which is qualitatively 
different from previously known nucleation mechanisms. 

\bigskip
\noindent
{\bf Results}

\noindent
We adopt the following Hamiltonian for the model,~\cite{Nishino_elastic}  
\begin{eqnarray}
{\cal H}_{0} = && \sum_{i=1}^{N} \frac{p_i^2}{2 m}
+ \sum_{i=1}^{N} V_i^{\rm intra}(r_i)   +  \sum_{i=1}^{N}
\frac{ \mbox{ \boldmath $P$}_i^2}{2 M}   \\ 
&& +  \sum_{\langle i,j\rangle
} V_{ij}^{\rm inter}( {\mbox{ \boldmath $X$}_i},{\mbox{ \boldmath $X$}_
j},r_i,r_j).   \nonumber
\end{eqnarray}
The first and second terms describe the motion of the intramolecular mode 
of the $i$th molecule. The radius of the molecule is $r_i$, and the conjugate 
momentum is $p_i$. The mass for the motion is $m$ (Supplymentary 2). 
The intramolecular potential energy $V_i^{\rm intra}(x_i)$ is shown by the solid curve in Fig.~\ref{Fig_model}a, where $x_i=r_i-r_{\rm LS}$. Here $r_{\rm LS}$ is the ideal radius of the LS molecule, and that of the HS molecule is $r_{\rm HS}=r_{\rm LS}+\Delta r$. $V_i^{\rm intra}(x_i)$ provides a symmetric vibration mode and induces changes of the molecular size. 
The third and fourth terms describe the ceneter-of-mass motion 
of the molecules ($\mbox{\boldmath $X$}_i$, $\mbox{ \boldmath $P$}_i$) with mass $M$. The intermolecular interaction $V_{ij}^{\rm inter}$  (see Methods) for nearest and next-nearest neighbors is a function not only of the coordinates $\mbox{\boldmath $X$}_i$ and $\mbox{\boldmath $X$}_j$, but also 
of the molecular radii $r_i$ and $r_j$. 
Although the interaction terms in this Hamiltonian appear only to be of short 
range, elastic interactions mediate the effect of the local lattice 
distortions over long distances.~\cite{Miyashita_elastic} 
We study relaxation and nucleation in this model at low temperatures 
by using a molecular dynamics method.~\cite{Nishino_elastic2} 
The critical temperature of the first-order transition of the model is $T_{\rm c} \simeq 0.9$, below (above) which the LS (HS) state is the equilibrium state.~\cite{Nishino_2010}  

We observe relaxation from the metastable HS state to the LS state 
at a low temperature ($T=0.2$) in approximately circular crystals on a square lattice.  
This temperature is much lower than the critical temperature $T_{\rm c}$.  
For the initial states of the relaxation (the metastable HS phase), we gave a 
set of velocities to all molecules according to the Maxwell-Boltzmann 
distribution  by using a random number sequence.  


Snapshots of the configuration during the course of a relaxation event 
are depicted in Fig.~\ref{Fig_model}d and Fig.~\ref{Fig_model}e, where the diameters of the circular crystals are 200 and 100 particles 
(denoted as $2R=200$ and $2R=100$), respectively. 
Figures~\ref{Fig_model}d1 and \ref{Fig_model}e1 show configurations when 
the HS fraction ($f_{\rm HS}$)~\cite{Nishino_elastic2} reaches the value $f_{\rm HS} \simeq 0.95$ for $2R=200$ and $2R=100$, respectively.  
Nucleation starts from one point along the circumference. 
The subsequent configurations are given in Figs~\ref
{Fig_model}d(2$-$6) and \ref{Fig_model}e(2$-$6) for $2R=200$ and $2R=100$, 
respectively. The corresponding values of 
$f_{\rm HS}$ are the same in both systems.  

As we show below, the configurations of Figs~\ref{Fig_model}d1 and \ref{Fig_model}e1 are those of the critical nucleus, and Figs~\ref{Fig_model}d(2$-$6) and Figs~\ref{Fig_model}e(2$-$6) correspond to deterministic growth of the LS droplet after the formation of the critical nucleus. 
It should be noted that the LS domain shapes can be well characterized by using the contact angle (wetting angle)~\cite{RICH97} of $\pi/2$. 
We checked the configurations for other relaxations from different initial 
conditions (different random number sequences for the molecular velocities) 
and found the same features of nucleation and growth. 
Here the size of the critical nucleus is found to be proportional to the 
system size. The shapes of the critical nuclei and also the following clusters 
are geometrically similar in systems of different sizes. 
This feature is qualitatively different from that of short-range interaction 
systems,~\cite{Rikvold} in which the critical droplet has a specific size 
independently of the system size. 

\begin{figure}
\centerline{\includegraphics[clip,width=5.4cm]{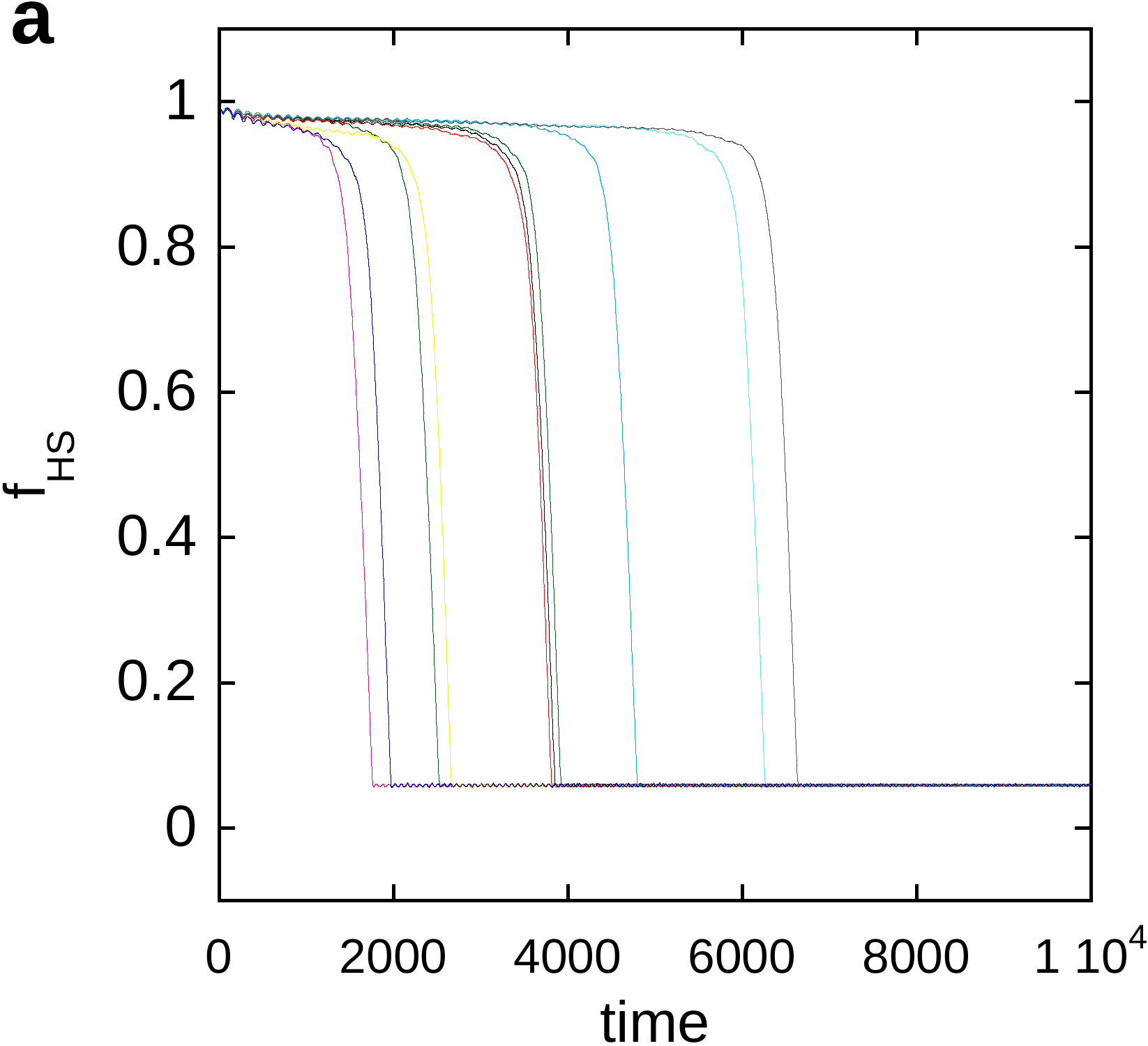} 
\includegraphics[clip,width=5.4cm]{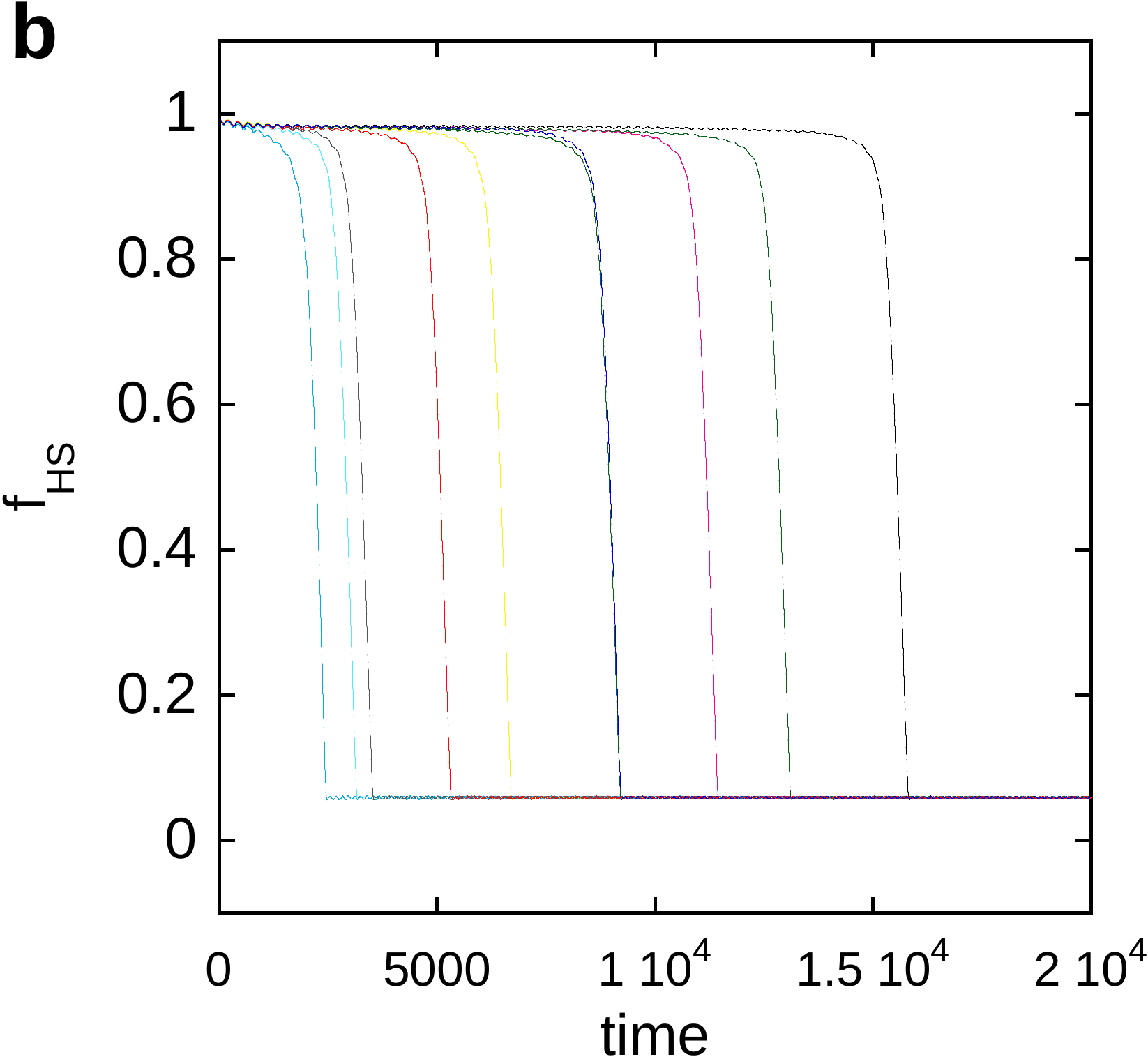} }

\vspace{1.4cm}
\centerline{\includegraphics[clip,width=12.0cm]{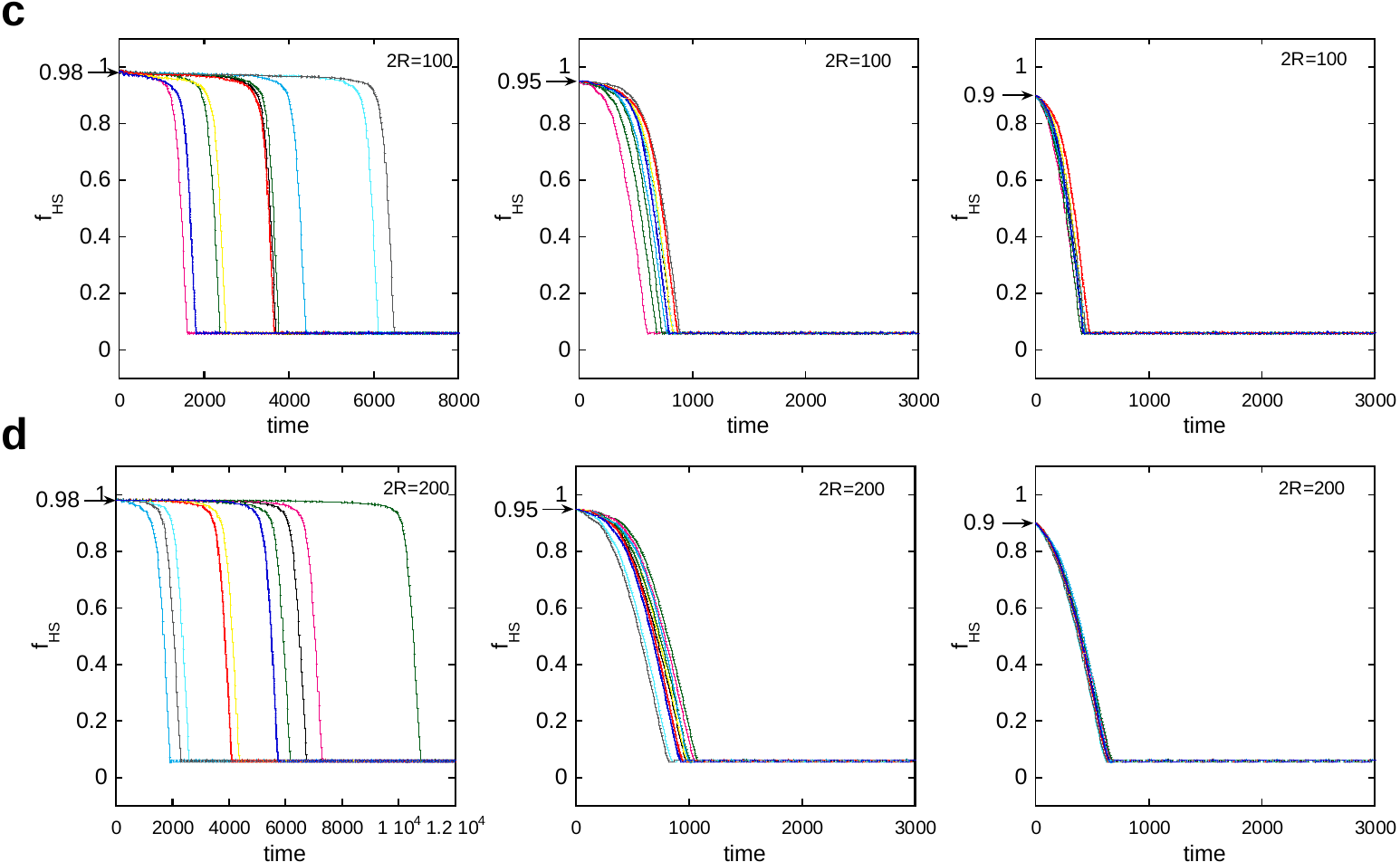} }
\caption{ {\bf Relaxation processes from the metastable HS phase.}
{\bf a}, {\bf b}, HS fraction ($f_{\rm HS}$) versus time at $T=0.2$ for 
({\bf a}) $2R=100$ and ({\bf b}) $2R=200$. The time of the collapse of the metastable state is widely distributed. 
{\bf c}, {\bf d}, The three panels of ({\bf c}) show the time dependence of $f_{\rm HS}$, starting from the time $t_{\rm tr}$, for $2R=100$ when $f_{\rm HS}^{\rm tr}$=0.98, 0.95, and 0.9 (from left to right). The three panels of ({\bf d}) correspond to the case of $2R=200$. 
}
\label{Fig_relax}
\end{figure}

In Figs~\ref{Fig_relax}a and \ref{Fig_relax}b the time dependence of 
$f_{\rm HS}$ is shown for systems with $2R=100$ and $2R=200$, respectively. 
Because a single nucleation event dominates the process, 
the escape time from the metastable state is random and 
governed by a Poisson process. 
However, once nucleation starts, the process is almost deterministic. 
The crossover between the stochastic and deterministic regimes determines 
the critical nucleus size and a threshold value of $f_{\rm HS}$. 
These are typical characteristics of barrier-crossing dynamics.

To capture this feature, we study the relaxation of $f_{\rm HS}$ after passing 
a given value of $f_{\rm HS}(\equiv f_{\rm HS}^{\rm tr})$. 
The passing time $t_{\rm tr}$ is defined as $f_{\rm HS}^{\rm tr}=f_{\rm HS}(t_{\rm tr}$). 
As mentioned above, the time evolutions after passing the threshold value, i.e., $f_{\rm HS}(t-t^{\rm tr})$ are expected to overlap. We plotted the data of 
$f_{\rm HS}(t-t^{\rm tr})$ for various trial values of $f_{\rm HS}^{\rm tr}$, and found $f_{\rm HS}(t^{\rm tr})\simeq 0.95$ gives the threshold as depicted in Fig.~\ref{Fig_relax}c. 
The same value is observed in both systems with 
$2R=100$ (Fig.~\ref{Fig_relax}c) and $2R=200$ (Fig.~\ref{Fig_relax}d), 
and we conclude that it is independent of the system size.

\bigskip
\noindent
{\bf Discussion}

\begin{figure}
\centerline{\includegraphics[clip,width=7.0cm]{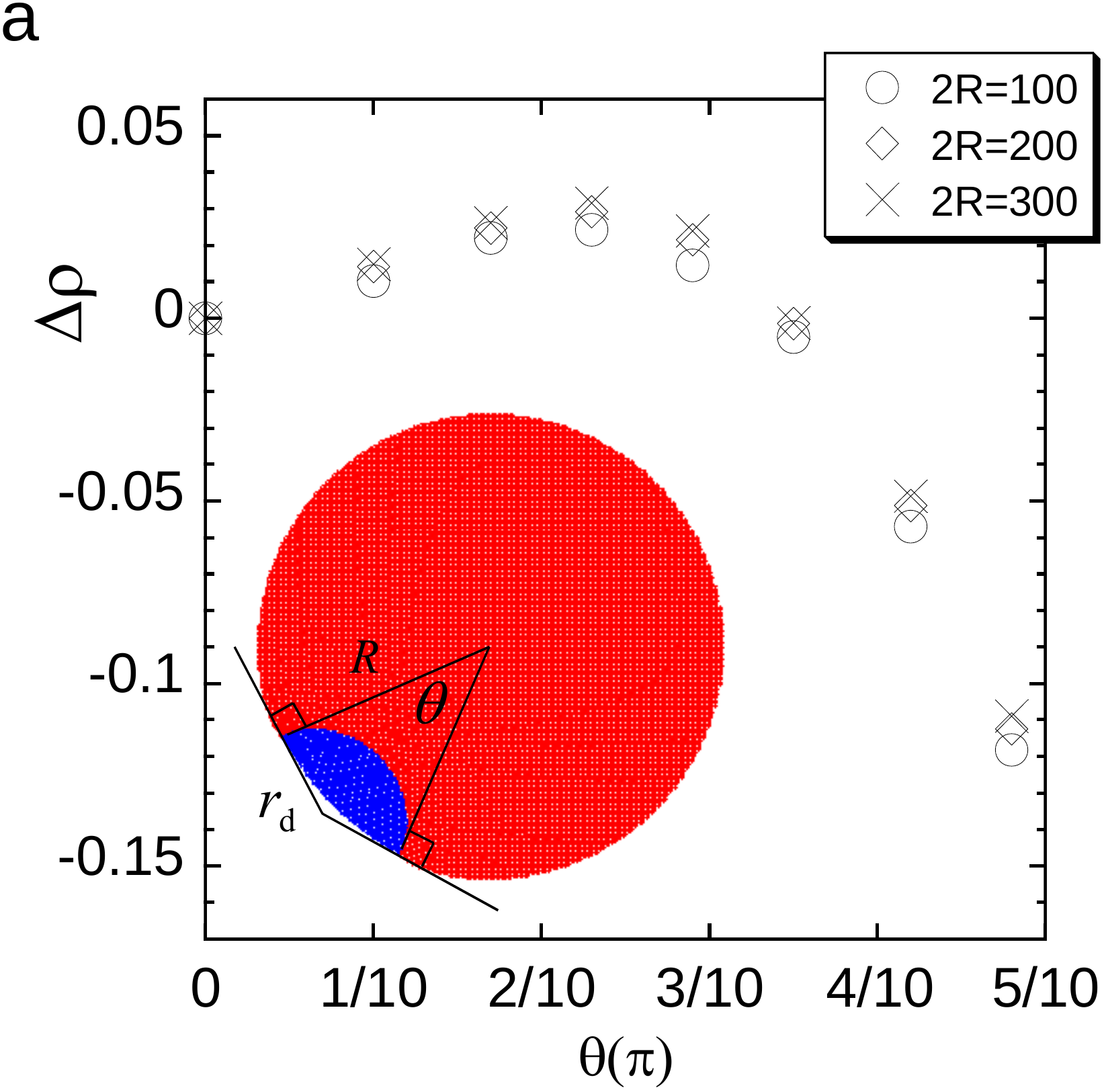} }

\vspace{0.8cm}
\centerline{\includegraphics[clip,width=6.0cm]{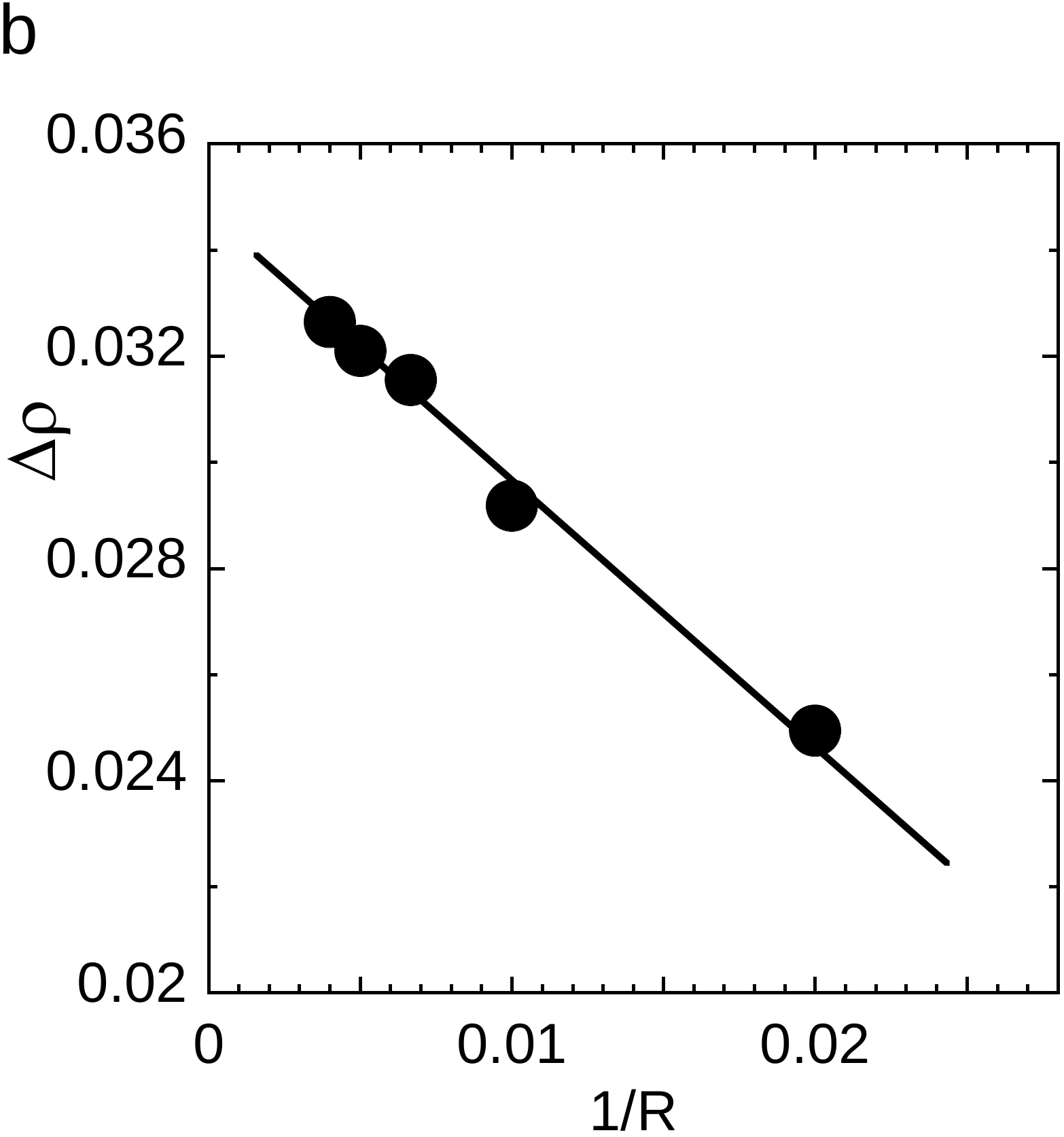} }
\caption{ {\bf Barrier-crossing of macroscopic nucleation.} 
{\bf a}, The excess energy density $\Delta \rho$ as a function of $\theta$ for the system sizes $2R=100,200$, and 300. The inset is the definition of $\theta$ and 
domain region is given for $\theta$ using the contact angle of $\pi/2$. 
For all $R$, the values of $f_{\rm HS}=$ 0.98, 0.97, 0.95, 0.92, 0.90, 
0.86, and 0.83 are given at $\theta / \pi=1/10$, 1.7/10, 2.3/10, 2.9/10, 3.5/10, 4.2/10, and 4.8/10 respectively. 
After $\theta/ \pi=4.8/10$, $\Delta \rho$ decreases monotonically until 
$\theta/ \pi=2$ (LS phase). 
{\bf b}, The dependence of $\Delta \rho$ on $1/R$ at $\theta=2.3\pi/10$ for $2R=100$, 200, 300, 400, and 500.
$\Delta \rho$ approaches a finite value as $R$ approaches infinity. 
 }
\label{Fig_theta-dep}
\end{figure}

\noindent
To examine the features of the critical nucleus and check the size dependence, 
we analyze the total potential energy of the system ($E_{\rm tot}=\sum V_i^{\rm intra}+V_{ij}^{\rm inter}$) as a function of the relative size of 
the LS domain. It is considered that the entropy effect is small enough 
compared to the energy barrier during the relaxation at this low $T$. 
As a parameter to characterize the domain size, we define $\theta$ as the 
central angle. With the contact angle of $\pi/2$, the domain region 
(lens-shaped part) is defined for any $\theta$  (Fig.~\ref{Fig_theta-dep}a), 
where the interface 
between the two phases is given by the circle of the radius 
$r_{\rm d}=R \tan (\theta/2)$, whose center is the crossing point of the two 
tangential lines. 
The value of $E_{\rm tot}$ for a given $\theta$ is obtained as follows.  
In the circle of the HS phase, we replace HS molecules in the lens-shaped part 
subtended by $\theta$ by LS molecules. Then we move all molecules slowly so as 
to reach the minimum total potential-energy state, and obtain the energy value 
of this stationary state. 
We define the energy density as $\rho=\frac{E_{\rm tot}}{N}$, where $N$ is the number of molecules in the system and $N \simeq \pi R^2$, 
and also the relative energy density: $\Delta \rho=\rho-\rho_{\theta=0}$ as the difference between $\rho$ of the stationary state and that of the complete HS phase ($\rho_{\theta=0}$).

We show $\Delta \rho$ as a function of $\theta$ for several system sizes ($2R$) in Fig. \ref{Fig_theta-dep}a. 
For small values of $\theta (\le \pi/10 $), $\Delta \rho$ is almost constant and then $\Delta \rho$ increases with $\theta$. In this region the cluster is expected to shrink in the relaxation process. Around $\theta=2.3\pi/10$, $\Delta \rho$ shows the maximum value and it decreases for larger $\theta$. 
It should be noted that at this $\theta$(=2.3$\pi$/10 ) $f_{\rm HS}$ is 
equal to 0.95, which agrees with the threshold value of $f_{\rm HS}$ in the analysis of the relaxation curves (Fig.\ref{Fig_relax}c and d).

When the droplet size exceeds the critical size, the domain expands. 
For different system sizes, this critical size of the droplet ($r_{\rm d}$) changes in proportion to 
the system syze ($R$). Namely, the critical angle exists, but {\em not} a 
specific critical size. This fact was demonstrated in Fig.~\ref{Fig_model}d 
and Fig.~\ref{Fig_model}e, i.e., the domain shape is almost the same for 
systems of different size. 
Thus we call this process ``macroscopic nucleation", and we believe  
that it should hold even in the bulk (continuum) limit. 
 We depict $\Delta \rho$ at $\theta=2.3\pi/10$ (peak position) as a function 
of $1/R$ in Fig.~\ref{Fig_theta-dep}b, and find the dependence: $\Delta \rho = 
\Delta \rho_0 -\frac{\rm const.}{R}$. The value $\Delta \rho_0 \approx 0.035$ 
is considered the value of the bulk limit. 

The behavior of ``macroscopic nucleation" is qualitatively different from that observed in short-range interaction systems. 
The bulk and surface contributions to the potential barrier cannot be 
distinguished in this long-range interaction system, which is similar to 
interface energies of binary alloys due to elasticity~\cite{Schulz}, 
and the elastic 
interactions suppress both bulk nucleation and multi-droplet nucleation 
at the boundary (Supplementary 1 and 3). 

In summary, we propose a new concept of ``macroscopic nucleation"   
for systems with long-range interactions. The domain formation exhibits 
geometric similarity for circular crystals of any size. 
This means that the size of the critical nucleus is proportional to the system 
size and macroscopic nucleation is realized. 
Recognition of this mechanism should give 
important insights for all systems in which local structural changes cause a 
distortion of the lattice. In addition to the spin-crossover type systems considered here, the mechanism should hold for martensitic and Jahn-Teller systems, etc.

\bigskip
\noindent
{\bf Methods}

\noindent
The role of the intermolecular potential is to release the local distortions due to the difference of the sizes of neighboring molecules. 
For this purpose, we adopt the following potential,~\cite{Nishino_elastic}   
$V_{ij}^{\rm inter}( {\mbox{ \boldmath $X$}_i},{\mbox{ \boldmath $X$}_j},
r_i,r_j)=f(d_{ij}-\Delta r)$, where $f(u) = D \left( e^{a' (u-u_0)} + e^{-b' (u-u_0)} \right)$. The variable $u_0$ is a constant such that $f(u)$ has its minimum at $u=0$ and $d_{ij}=| {\mbox{ \boldmath $X$}_i} - {\mbox{ \boldmath $X$}_j} |- (r_i+r_j)$. For nearest neighbors, $\Delta r=0$, $a'=0.5$ and $b'=1.0$ are set, and the energy minimum is realized when the neighbors have the same size. For next-nearest neighbors, $\Delta r= 2(\sqrt{2}-1) \bar{r}$ 
with $\bar{r}=(r_{\rm LS} + r_{\rm HS}) /2$, $a'=0.1$ and $b'=0.2$ are set. 
This provides a small force sufficient to ensure the stability of the crystal structure (this is specific to coordination $z=4$). The parameter $D$ associated with the strength of the intermolecular interaction was set to $D=20$, which is strong enough to cause a first-order phase transition. 
Here the type of the potential function is not so important, and 
the basic mechanism of macroscopic nucleation is universal for other types 
of intermolecular potentials (harmonic or anharmonic potentials). 
The other parameters were set as $r_{\rm HS}=9$, $\Delta r=1$, and $m=M=1$ 
(Supplementary 2).  
Molecular dynamics simulations were performed using a 
Nos{\'e}-Hoover thermostat.~\cite{Nishino_elastic2} 
With this method, the timescale of the simulation is influenced by the
thermostat parameters. Here, we used this effect to our advantage to
perform the simulations for large systems in a computationally feasible
time. Although we sacrifice the ability to measure nucleation times, which
we expect to increase dramatically with system size, our method allows us
to observe the scale-invariant spatial structure of the nucleation process
in systems of very different sizes, as shown in Figs~\ref{Fig_model}d and \ref{Fig_model}e. The
qualitative feature of nucleation in a stochastic Poisson process,
followed by deterministic growth shown in Fig.~\ref{Fig_relax}, is also preserved.

\bigskip
\noindent
{\bf Acknowledgements}

\noindent
The authors thank I. Chiorescu for a useful comment. 
The present work was supported by Grant-in-Aid for Scientific Research on Priority Areas (17071011) and for Scientific Research C (23540381), 
and by the Next Generation Super Computer Project, 
Nanoscience Program from MEXT of Japan. 
CE acknowledges a 185/2010 Romanian CNCS Young Researchers Grant. 
PAR acknowledges US NSF Grants No. DMR-0802288 and DMR-1104829. 
The numerical calculations were supported by the supercomputer center of
ISSP of the University of Tokyo. 

\bigskip
\noindent
{\bf Author Contributions}

\noindent
M. N and S. M. planned this subject and 
M. N. obtained most of the data by performing computational simulations.  
M. N., C. E., S. M., P. A. R., K. B., and F. V. contributed to analyses and discussions of the results and also to preparation of the manuscript. 

\bigskip
\noindent
{\bf Additional information}

\noindent
Supplementary Information accompanies this paper at 
http://www.nature.com/scientificreports. Competing financial interests: The authors declare no competing financial interests. 
Correspondence and requests for materials should be addressed to M. N. (nishino.masamichi@nims.go.jp).

\end{document}